\documentclass[twocolumn]{aastex62}

\newcommand\Alfven{Alfv\'en \,}
\newcommand{\be}{\begin{equation}}
\newcommand{\ee}{\end{equation}}

\usepackage{epsfig}
\usepackage{epstopdf}
\usepackage{graphicx}
\usepackage{amsmath}
\usepackage{appendix}

\submitjournal{ApJ}
\shorttitle{Towards a Full MHD Jet Model}
\shortauthors{Huang et al.}

\begin{document}
\title{Towards a Full MHD Jet Model of Spinning Black Holes--I: Framework and a split monopole example}

\author{Lei Huang}
\affiliation{Key Laboratory for Research in Galaxies and Cosmology, Shanghai Astronomical Observatory, \\Chinese Academy of Sciences, Shanghai, 200030, China}
\email{muduri@shao.ac.cn}

\author{Zhen Pan}
\affiliation{Perimeter Institute for Theoretical Physics, 31 Caroline St N, Waterloo, Ontario, N2L2Y5, Canada}
\email{zpan@perimeterinstitute.ca}

\author{Cong Yu}
\affiliation{School of Physics and Astronomy, Sun Yat-sen University, Zhuhai 519082, China}
\email{yucong@mail.sysu.edu.cn}

\begin{abstract}
In this paper, we construct a framework for investigating magnetohydrodynamical jet structure of spinning black holes (BHs), where electromagnetic fields and fluid motion are governed by the Grad-Shafranov equation and the Bernoulli equation, respectively. Assuming steady and axisymmetric jet structure, we can self-consistently obtain electromagnetic fields, fluid energy density and velocity within the jet, given proper plasma loading and boundary conditions. Specifically, we structure the two coupled governing equations as two eigenvalue problems, and develop full numerical techniques for solving them. As an example, we explicitly solve the governing equations for the split monopole magnetic field configuration and  simplified plasma loading on the stagnation surface where the poloidal fluid velocity vanishes. As expected, we find the rotation of magnetic field lines is dragged down by fluid inertia, and the fluid as a whole does not contribute to energy extraction from the central BH, i.e., the magnetic Penrose process is not working. However, if we decompose the charged fluid as two
oppositely charged components, we find the magnetic Penrose process does work for one of the two components when the plasma loading is low enough.
\end{abstract}


\section{Introduction} \label{sec:intro}

\begin{figure}[h!]
\centering
\includegraphics[scale=1]{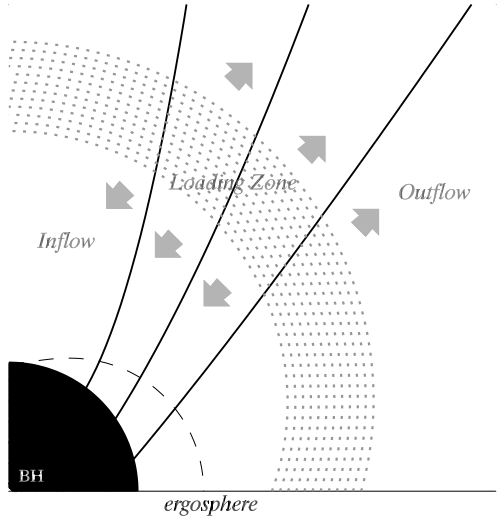}
\caption{A cartoon picture of our MHD model of BH jets, 
where magnetic field lines (solid lines) cross the ergosphere (dashed line) and the event horizon. 
There is an extending plasma loading zone (shaded region) near the central BH, where particles are injected. 
The inflow and outflow pattern naturally forms under the mutual influence of the central BH and the EM fields. \label{fig:cartoon}}
\end{figure}

Relativistic jets launched by accreting black holes (BHs) play an essential role in several energetic astrophysical phenomena,
including stellar-mass BH X-ray binaries, active galactic nuclei and possibly gamma-ray bursts.
After decades of debates among  astrophysical communities, many open questions concerning the nature of the BH jets still remain to be answered \citep[see e.g.,][for reviews]{Meier01, Blandford18}. To name a few of the most fundamental ones: what are the central engines of the jets; how the fluid within the jets is accelerated to the relativistic speed; how the jets are collimated.

The Blandford-Znajek (BZ) mechanism \citep{BZ77,Znajek77}, which describes an electromagnetic (EM) process of extracting the rotation energy of the central BH in the form of Poynting flux, is believed to be the most promising candidate for the central engines
of the BH jets. For understanding the jets powered by the BZ mechanism, one needs to study magnetohydrodynamic (MHD) process in the Kerr spacetime, where the EM fields and the fluid motion are coupled in a complicate way. Therefore people treat some components of the MHD process as dynamical variables with other components prescribed in many previous studies on this subject.
For studying the EM fields of the jet, force-free electromagnetodynamics (FFE) is a convenient assumption, where
the fluid energy is ignored and the EM fields are self-contained \citep[e.g.,][]{Tanabe08, Kom01, Kom02, Kom04a, 
Kom04b, Kom05, Kom07,Beskin13, Contop13,Nathan14, Gralla14,Gralla15,Gralla16,Pan14,Pan15,Pan15b,Pan16,Pan17,Pan18, 
Yang14, Yang15, East18, Mahlmann18}.
For studying the fluid motion within the jet, people usually treat the fluid as test fluid in prescribed EM fields \citep[e.g.,][]{Takahashi90, Beskin06, Globus13,Globus14,Pu12, Pu15}. There have also been some full MHD attempts where only
outflow pattern and jet structure in weak gravity regime are addressed \citep[see e.g.,][for self-similar outflow solutions in pseudo-potential]{Polko13, Polko14,Cecco18} 
and \citep[see e.g.,][for outflow solutions in Minkowski spacetime]{Beskin98, Beskin00,Lyub09, Tchek09, Beskin10,Beskin17}.
For understanding the physics of BH accretion systems, general relativistic MHD (GRMHD) simulation has been another powerful tool 
in the past two decades, in which full MHD equations in curved spacetime are solved. 
Nevertheless,  GRMHD codes tend to become unstable to a vacuum and therefore a matter density floor is usually introduced 
\citep[e.g.,][]{Gammie03, Shibata05, Porth17}, which may obscure our understanding of plasma loading and the flow within the jet. 

Besides all the theoretical explorations summarized above,  substantial progress
in spatial resolution has been made on the observation side. Especially the Event Horizon Telescope (EHT) 
\citep[e.g.,][]{Doel08,Doel12,Ricarte15,EHT19I,EHT19V} is expected 
to resolve the structure of supermassive BHs nearby (Sgr A$^*$ and  M87) down to horizon scales.
It is promising to unveil the physical nature of the jets in these systems if the coming EHT observations can 
be correctly deciphered.  This motivates us to construct a full GRMHD jet model, considering that 
all the previous studies are confined by different limitations.

In this paper, we aim to construct a GRMHD framework for investigating the structure of steady and axisymmetric jets 
of spinning BHs, in which the EM fields and fluid motion are self-consistently taken care of. 
A cartoon picture in Fig.~\ref{fig:cartoon} is shown to illustrate the major elements of our jet model:
a central BH, EM fields, a plasma loading zone, inflow and outflow. 
The magnetic field lines penetrate the event horizon of a spinning BH and 
extract the rotation energy from the BH in the form of Poynting flux.
Quantifying plasma loading within the BH jet is also a complicate problem considering 
the rich plasma sources, including the accretion flow centered on the equatorial plane, 
pair production inside the jet \citep{Lev11, Brod2015, Hiro16, Chen18}
and neutrino pair annihilation from an extremely hot accretion flow \citep[see e.g.][]{Pop99, Narayan01}. 
In our jet model, we do not deal with these detailed processes. For convenience, we 
introduce a plasma loading zone where plasma is injected and prescribe the loading function, i.e., particle number flux per magnetic flux  $\eta(r,\theta)$.  Under the mutual influence of the central BH and the EM fields, the inflow and outflow pattern naturally forms.
In  summary: we aim to construct a framework for investigating MHD jet structure of spinning BHs, in which the EM fields and the fluid motion are self-consistently obtained given  proper boundary conditions and a proper plasma loading function $\eta(r,\theta)$.

This paper is organized as follows. In Section~\ref{sec:setup}, we summarize some basic equations and assumptions to be used in this paper.  We derive the two governing equations: the Bernoulli equation and the MHD Grad-Shafranov (GS) equation in Section~\ref{sec:Bern} and Section~\ref{sec:GS}, respectively. We detail the numerical techniques
for solving the governing equations in Section~\ref{sec:eg}. 
The numerical solutions of MHD jet structure with split monopole magnetic field configuration are presented in Section~\ref{sec:results}. 
Summary and discussion are given in Section~\ref{sec:summary}. 
For reference, we place some  details for deriving the governing equations in Appendix \ref{sec:D_der} and \ref{sec:GS_der}.
Throughout this paper, we use the geometrical units $c=G=M=1$, where $M$ is the mass of the central BH.

\section{Basic Setting Up} \label{sec:setup}

The background  Kerr metric is written in the Boyer-Lindquist coordinates as follows,
\begin{eqnarray}
	{\rm d}s^2&=& g_{tt} {\rm d}t^2 + 2g_{t\phi}{\rm d}t {\rm d}\phi + g_{\phi\phi} {\rm d}\phi^2 + g_{rr} {\rm d}r^2 + g_{\theta\theta} {\rm d}\theta^2  \nonumber\\
	&\ &  \nonumber\\
	&=& \left( \frac{2Mr}{\Sigma}-1 \right) {\rm d}t^2 - 2\ \frac{2Mar\sin^2\theta}{\Sigma} {\rm d}t {\rm d}\phi  \nonumber\\
	&\ & + \frac{\beta\sin^2\theta}{\Sigma} {\rm d}\phi^2 + \frac{\Sigma}{\Delta} {\rm d}r^2 + \Sigma {\rm d}\theta^2\ ,
\end{eqnarray}
where $a$ and $M$ are the BH spin and mass, respectively, $\Sigma=r^2+a^2\cos^2\theta$, $\Delta=r^2-2Mr+a^2$,  $\beta =(r^2+a^2)^2-a^2\Delta\sin^2\theta$ and the square root of the determinant $\sqrt{-g}=\Sigma\sin\theta$.

We investigate the structure of a steady and axisymmetric BH jet and  we assume the plasma within the jet is perfectly conducting, i.e., $\partial_t = \partial_\phi = 0$ and $\mathbf{E} \cdot \mathbf{B}=0$, where $\mathbf{E}$ and 
$\mathbf{B}$ are the electric and the magnetic fields, respectively. Then all the non-vanishing components of Maxwell tensor are expressed as follows \citep[see e.g.,][]{Pan14}
\begin{equation}
\label{eq:Maxwell}
\begin{aligned}
    F_{r\phi} &= -F_{\phi r} =\Psi_{,r}  \ ,   &F_{\theta\phi} &= -F_{\phi\theta} = \Psi_{,\theta} \ , \\
    F_{tr} &= -F_{rt} =\Omega\Psi_{,r}  \ ,    &F_{t\theta} &= -F_{\theta t} = \Omega  \Psi_{,\theta} \ , \\
    F_{r\theta} &= -F_{\theta r} = - \frac{\Sigma }{\Delta\sin\theta} I \ ,
\end{aligned}
\end{equation}
where $\Psi = \Psi(r,\theta)$ is the magnetic flux and $\Omega = \Omega(\Psi)$ is the angular velocity of magnetic field lines. 
For convenience, we have defined poloidal electric current $I(r,\theta) \equiv \sqrt{-g} F^{\theta r}$. Therefore,
the EM  fields are completely determined by three quantities: $\{\Psi(r,\theta), \Omega(\Psi), I(r,\theta)\}$.

Before proceeding on, it is useful to define a few conserved quantities.
From the perfectly conducting condition $F_{\mu\nu} u^\nu = 0$, we find different components
of fluid velocity are related by
\begin{equation}
 \frac{ u^r}{\Psi_{,\theta}} = - \frac{u^\theta}{\Psi_{,r}}
	= \frac{(u^\phi-\Omega u^t)}{F_{r\theta}}\ ,
\end{equation}
from which we can define the particle number flux per magnetic flux
\begin{equation}\label{eq:eta}
\begin{aligned}
 \eta&\equiv\frac{\sqrt{-g} n u^r}{\Psi_{,\theta}} = - \frac{\sqrt{-g} n u^\theta}{\Psi_{,r}} \\
	&= \frac{\sqrt{-g} n(u^\phi-\Omega u^t)}{F_{r\theta}} \ .
\end{aligned}
\end{equation}
From the energy-momentum tensor $T^{\mu\nu} = T^{\mu\nu}_{\rm EM} + 	T^{\mu\nu}_{\rm MT}$, 
where the EM part and the matter (MT) part are 
\begin{equation}
\label{eq:energy_tensor}
\begin{aligned}
	T^{\mu\nu}_{\rm EM}&= \frac{1}{4\pi} \left( F^{\mu\rho}F_{\ \ \rho}^{\nu} - \frac{1}{4}g^{\mu\nu}F_{\alpha\beta}F^{\alpha\beta} \right)\ , \\
	T^{\mu\nu}_{\rm MT}&= \rho u^\mu u^\nu = nm  u^\mu u^\nu \ ,
\end{aligned}
\end{equation}
we define total energy per particle $E$ and total angular moment $L$ per particle as follows,
\begin{equation}\label{eq:EandL}
\begin{aligned}
	E&\equiv E_{\rm MT}+E_{\rm EM}= -mu_t +\frac{\Omega I}{4\pi\eta}\ ,  \\
	L&\equiv L_{\rm MT}+L_{\rm EM} = mu_\phi +\frac{I}{4\pi\eta}\ ,
\end{aligned}
\end{equation}
where  $\rho$, $n$ and $m$ are the  proper energy density, the proper number density and the particle rest mass, respectively;
and we have assumed cold plasma.

Now let us examine the conservation property of these quantities along magnetic field lines.
For this purpose, we define derivative along field lines
\begin{equation}
    D^\parallel_\Psi \equiv \frac{1}{\sqrt{-g}}(\Psi_{,\theta}\partial_r -  \Psi_{,r}\partial_\theta)\ ,
\end{equation}
and it is straightforward to obtain (see Appendix \ref{sec:D_der})
\begin{equation} \label{eq:D_eta}
	D^\parallel_\Psi\eta = (nu^\mu)_{;\mu} \ ,
\end{equation}
i.e., $D^\parallel_\Psi\eta$ quantifies the plasma loading rate. In general, we can write the
energy-momentum conservation as 
\begin{equation}\label{eq:smu}
    T^{\mu\nu}_{\phantom{xy};\nu} = S^\mu,
\end{equation}
where the source term $S^\mu$ comes from plasma loading. As a simple example, we assume  
 $S^\mu = (D^\parallel_\Psi\eta) mu^\mu$ in this paper,
i.e., the source term is contributed by the kinetic energy of newly loaded plasma.
With a few steps of calculation as detailed in Appendix \ref{sec:D_der}, we obtain
\be\label{eq:D_etaEL}
\begin{aligned}
	D^\parallel_\Psi(\eta E)&= (D^\parallel_\Psi\eta)(-mu_t)\ ,\\
	D^\parallel_\Psi(\eta L)&= (D^\parallel_\Psi\eta)(mu_\phi)\ ,
\end{aligned}
\ee
where $\eta E$ and $\eta L$ are the energy flux per magnetic flux and angular momentum flux per magnetic flux, respectively.
\emph{Outside} the plasma loading zone where there is no particle injection,
the particle number conservation reads as
\begin{eqnarray}
	\left(n u^\mu\right)_{;\mu}&=&0\ ,
\end{eqnarray}
therefore $\eta, E, L$ are conserved along field lines,
i.e., $\eta=\eta(\Psi), E = E(\Psi), L=L(\Psi)$.

In summary: with assumptions of steady and axisymmetric jet structure and perfectly conducting plasma within the jet, we have obtained one conserved quantity $\Omega(\Psi)$
and three ``quasi-conserved"  quantities $\{\eta(\Psi), E(\Psi), L(\Psi)\}$
which are only conserved \emph{outside} the plasma loading zone.

\section{Bernoulli equation} \label{sec:Bern}

From the normalization condition $u^\mu u_\mu=-1$ and Eqs.~(\ref{eq:eta},\ref{eq:EandL}), we obtain the relativistic Bernoulli equation
\begin{eqnarray}\label{eq:Bern}
	\mathcal{F}(u) = u_p^2+1-\left(\frac{E}{m}\right)^2 U_g(r,\theta) = 0\ ,\end{eqnarray}
where $u_p^2 \equiv u^Au_A$ with the dummy index $A=\{r,\theta\}$.
In the Kerr space-time,  the characteristic function $U_g$ is writen as \citep{Camen86a,Camen86b,Camen87,Takahashi90,Fendt01,Fendt04,Levinson06,Pu15}
\begin{eqnarray}
\label{Ug}
	U_g(r,\theta)&=& \frac{k_0k_2-2k_2\mathcal{M}^2-k_4\mathcal{M}^4}{(\mathcal{M}^2-k_0)^2}\ ,
\end{eqnarray}
where
\begin{eqnarray}
\label{Ugk}
	k_0&=& -[g_{tt}+2g_{t\phi}\Omega+g_{\phi\phi}\Omega^2]\ ,\nonumber\\
	k_2&=& \left[ 1-\Omega(E/L)^{-1} \right]^2\ ,\nonumber\\
	k_4&=& \frac{\left[ g_{tt}(E/L)^{-2} + 2g_{t\phi}(E/L)^{-1} +g_{\phi\phi} \right]}{g_{tt}g_{\phi\phi}-g_{t\phi}^2}\ ,
\end{eqnarray}
and the \Alfven Mach number $\mathcal{M}$ is given by
\begin{equation}\label{eq:mach}
    \mathcal{M}^2=\frac{4\pi m\eta^2}{n} = 4\pi mn\frac{u_p^2}{B_p^2} = 4\pi m\eta\frac{u_p}{B_p},
\end{equation}
with the poloidal magnetic field $B_p$ defined by
\be
(\sqrt{-g} B_p)^2 = g_{rr} (\Psi_{,\theta})^2 + g_{\theta\theta} (\Psi_{,r})^2  \ .
\ee

Several characteristic surfaces can be defined according to the critical points of the flow velocity
\citep[see e.g., ][for details]{Michel69,Michel82,Camen86a,Camen86b, Takahashi90, Beskin09}.
The light surface (LS) is defined by where the rotation velocity of field lines approaches light speed and where
particles are forbidden to corotate with the field lines,
\be
k_0  \big|_{r=r_{\rm LS}} = 0 \ .
\ee
The \Alfven surface is defined by where the denominator of characteristic function $U_g(r,\theta)$ vanishes, i.e.,
\be
    - k_0 + \mathcal{M}^2 \big|_{r=r_A} = 0 \ .
\ee
On the \Alfven surface, we find 
\be 
\frac{E}{L}= - \frac{g_{tt} + g_{t\phi}\Omega }{g_{t\phi} + g_{\phi\phi}\Omega}\Bigg|_{r=r_A}\ , 
\ee 
where we have used Eqs.(\ref{eq:Bern}-\ref{Ugk}).
The stagnation surface where $u_p = 0$ is determined by
\be\label{eq:stag}
D^\parallel_\Psi k_0 \big|_{r=r_*}=0 \ .
\ee
The fast magnetosonic (FM) surface and  slow magnetosonic (SM) surface are defined by where the denominator of $D_\Psi^\parallel u_p$ vanishes. In the cold plasma limit, the SM surface coincides with the stagnation surface.
On the stagnation surface, where both $u_p$ and $\mathcal{M}$ vanish, we find
\be \label{eq:Estag}
\left(\frac{E}{m} \right)^2 = \frac{k_0}{k_2}\Bigg|_{r=r_*}\ ,
\ee
where we have used Eqs.(\ref{eq:Bern}, \ref{Ug}). 

Plugging Eq.(\ref{eq:eta}) into Eq.(\ref{eq:Bern}), we find that the Bernoulli equation is a polynomial equation of
fourth order in $u_p$ with to-be-determined eigenvalue $E/L$ (or equivalently the location of the \Alfven surface $r_A$), given prescribed angular velocity $\Omega$ and particle number flux per field line $\eta(r,\theta)$ \citep[see e.g.,][]{Camen86a,Camen86b, Takahashi90, Fendt01, Pu12, Pu15}.

\subsection{Single Loading Surface}

As a first step towards a full MHD jet solution, we mathematically idealize the plasma loading zone as a single surface, and we choose the stagnation surface (Eq.~(\ref{eq:stag})) as the plasma loading surface \citep[see e.g.,][ for a detailed gap model]{Brod2015} in this paper. 
To define the plasma loading for both inflow and outflow, 
we introduce a pair of dimensionless magnetization parameters on the loading surface,
\be\label{sigM}
	\sigma_*^{\rm in;out} =\frac{B_{p,*}}{4\pi m |\eta|_{\rm in;out}}\ ,
\ee
where $B_{p,*}$ is the poloidal field on the loading surface.
In this way, the particle number flux per magnetic flux $\eta$ is completely determined by $\sigma_*^{\rm in;out} $, recalling that
$\eta$ is a conserved quantity along field lines outside the loading zone. Note that $\eta_{\rm in} < 0$ and
$\eta_{\rm out} > 0$, therefore there is a jump in $\eta$ at the loading surface, i.e., 
$D_\Psi^\parallel \eta \propto \delta(r-r_*)$.

Using Eq.(\ref{eq:Estag}), the Bernoulli equation (\ref{eq:Bern}) can be rewritten into a fourth-order polynomial equation as
\begin{eqnarray} \label{Bern2}
	\sum_{i=0}^4A_i u_p^i&=&0\ ,
\end{eqnarray}
where the coefficients $A_i$ are functions expressed by
\begin{equation}
\begin{aligned}
	A_4&= \frac{1}{\sigma_*^2}\frac{B_{p,*}^2}{B_p^2}\ ,\\
	A_3&= -\frac{2k_0}{\sigma_*}\frac{B_{p,*}}{B_p}\ ,\\
	A_2&= k_0^2 + \left(1 + \frac{k_{0,*}}{k_{2,*}} k_4\right) \frac{1}{\sigma_*^2}\frac{B_{p,*}^2}{B_p^2}\ ,\\
	A_1&= \left(-k_0 + \frac{k_{0,*}}{k_{2,*}} k_2\right) \frac{2}{\sigma_*}\frac{B_{p,*}}{B_p}\ ,\\
	A_0&= k_0^2 - \frac{k_{0,*}}{k_{2,*}} k_0k_2\ .\\
\end{aligned}
\end{equation}

As explored in several previous studies \citep[e.g.,][]{Takahashi90, Pu15},
solving the Bernoulli equation above is in fact an eigenvalue problem, where $(E/L)_{\rm in}$ is the to-be-determined eigenvalue ensuring inflow smoothly cross the FM surface, while $(E/L)_{\rm out}$ is given by the match condition on the loading surface. Eq.(\ref{eq:D_etaEL}) provides  conditions connecting the inflow and the outflow for the single surface loading,
\begin{eqnarray}\label{eq:deltaEL}
	\delta(\eta E)&=& m(\delta\eta)(-u_t)_*\ ,\nonumber\\
	\delta(\eta L)&=& m(\delta\eta)(u_\phi)_*\ ,
\end{eqnarray}
which give the match condition
\begin{eqnarray}\label{eq:EOLout}
	(E/L)_{\rm out}&=& \frac{(\eta E)_{\rm out}}{(\eta L)_{\rm out}} =\frac{(\eta E)_{\rm in} + m(\delta\eta)(-u_t)_*}{(\eta L)_{\rm in} + m(\delta\eta)(u_\phi)_*}\ ,
\end{eqnarray}
where $\delta\eta\equiv\eta_{\rm out}-\eta_{\rm in}$, and we have used the fact that $D_\Psi^\parallel\eta$ is 
a $\delta$-function centered on the loading surface in deriving Eq.~(\ref{eq:deltaEL}).
It is straightforward to see Eq.~(\ref{eq:deltaEL}) guarantees a same jump in the total energy
flux and its matter component, therefore the Poynting flux (and all the EM field components) is continuous across the loading surface. 

As along as the Bernoulli equation is solved, i.e., both the eigenvalues $(E/L)_{\rm in, out}$ and the poloidal velocity field $u_p$ are obtained,
$u^r$ and $u^\theta$ is obtained via Eq.(\ref{eq:eta}) and $u_p^2=u^Au_A$,
while $u_t$ and $u_\phi$ are obtained via relation $m(u_t+\Omega u_\phi) = -(E-\Omega L)$ and the normalization condition $u\cdot u=-1$.

Before delving into the details of numerically solving the Bernoulli equation, we can now give an estimate of the eigenvalues. Combining the definitions of $E$ and $L$ Eqs.(\ref{eq:EandL}) with Eq.(\ref{eq:Bern}), we find
\be
(u_t+\Omega u_\phi)_*=-\sqrt{k_{0,*}},
\ee
plugging which back into Eqs.(\ref{eq:EandL}), we obtain
\begin{eqnarray}
\label{EOL}
	(E/L)_{\rm in}&=&\Omega + \frac{m\eta_{\rm in}\sqrt{k_{0,*}}}{(\eta L)_{\rm in}}\ <\Omega\ ,\nonumber\\
	(E/L)_{\rm out}&=&\Omega + \frac{m\eta_{\rm out}\sqrt{k_{0,*}}}{(\eta L)_{\rm out}}\ >\Omega\ ,
\end{eqnarray}
which imply $E/L = \Omega [1 + O(\sigma_*^{-1})]$ and we have used the fact $\eta_{\rm in}<0$ and 
$\eta_{\rm out}>0$. 

\section{MHD Grad-Shafranov equation} \label{sec:GS}
With the aid of the Maxwell's equation
\be
	 {F^{\mu\nu}}_{;\nu} = 4\pi j^\mu \ ,
\ee
the trans-field component of the energy conservation equation (\ref{eq:smu}) is written as \footnote{Eq.~(\ref{eqn_MHD}) 
only holds for the specific choice of source function $S^\mu = (nu^\nu)_{;\nu} mu^\mu$.} 
\be\label{eqn_MHD}
	\frac{F^A_{\ \phi}}{F_{C\phi}F^C_{\ \phi}} (mn u^\nu u_{A;\nu}-F_{A\nu}j^\nu)=0\ ,
\ee
where we have used Eq.~(\ref{eq:D_eta}) and the source function $S^\mu$. 
The repeated Latin letters $A$ and $C$ run over the poloidal coordinates $r$ and $\theta$ only 
\citep{Nitta91,Beskin93,Beskin97}.
This is known as the MHD GS equation, with the $1^{\rm st}$ and $2^{\rm nd}$ terms
in the bracket being the fluid acceleration and the electromagnetic force, respectively.

After some tedious derivation (see Appendix \ref{sec:GS_der}), we write the full MHD GS equation in a compact form
\begin{eqnarray}
\label{GS_MHD}
	&\ &\mathcal{L}\Psi=\mathcal{S}_{\rm EM}+\mathcal{S}_{\rm MT}\ .
\end{eqnarray}
Here $\mathcal{L}$ is a differential operator defined by
\begin{eqnarray}
\label{GS_MHD_L}
	\mathcal{L}\Psi&& = \left[\Psi_{,rr} + \frac{\sin^2\theta}{\Delta} \Psi_{,\mu\mu} \right]\ \mathcal{A}(r,\theta;\Omega)   \nonumber\\
	&\ & + \left[ \Psi_{,r} \partial^\Omega_r  + \frac{\sin^2\theta}{\Delta} \Psi_{,\mu} \partial^\Omega_\mu \right] \ \mathcal{A}(r,\theta;\Omega)  \nonumber\\
	&\ & + \frac{1}{2} \left[ (\Psi_{,r})^2 + \frac{\sin^2\theta}{\Delta} (\Psi_{,\mu})^2 \right] D_\Psi^\perp\Omega\ \partial_\Omega \mathcal{A}(r,\theta;\Omega)  \nonumber\\
	&\ & - \left[ (\Psi_{,r})^2 + \frac{\sin^2\theta}{\Delta} (\Psi_{,\mu})^2 \right] \frac{D^\perp_\Psi\eta}{\eta}\ \mathcal{M}^2(r,\theta)\ ,\nonumber\\
\end{eqnarray}
where  $\mu=\cos\theta$, $\mathcal{A}(r,\theta;\Omega)=-k_0(r,\theta;\Omega)+\mathcal{M}^2(r,\theta)$,
and we have defined $\partial^\Omega_A(A=r,\mu)$ as the partial derivative with respect to coordinate $A$ with $\Omega$ fixed, $\partial_\Omega$ as the derivative with respect to $\Omega$, $D_\Psi^\perp$ as the derivative perpendicular to field lines
\begin{eqnarray}
	D_\Psi^\perp&\equiv& \frac{F^A_{\ \phi}\partial_A}{F_{C\phi}F^C_{\ \phi}}\ ,
\end{eqnarray}
which is equivalent to the ordinary derivative $d/d\Psi$ when acting on functions of $\Psi$.
The two source terms are
\be
\label{GS_MHD_S}
\begin{aligned}
	 \mathcal{S}_{\rm EM} &=\frac{\Sigma}{\Delta} I D^\perp_\Psi I\ , \\
     \mathcal{S}_{\rm MT} &=-4\pi\Sigma\sin^2\theta mn(u^tD_\Psi^\perp u_t + u^\phi D_\Psi^\perp u_\phi)\ ,
\end{aligned}
\ee
where $I=4\pi(\eta L-\eta m u_\phi)$ [see Eq.(\ref{eq:EandL})]. 

In the FFE limit, $\mathcal{M}^2=0$, $\mathcal{S}_{\rm MT}=0$, and the  GS equation reduces to \citep{Pan17}
\begin{eqnarray}
\label{GS_FFE}
	&\ &\mathcal{L}\Psi = \mathcal{S}_{\rm EM}\ \ .
\end{eqnarray}
The FFE solutions $\{\Psi|_{\rm FFE}, \Omega|_{\rm FFE}, (\eta L)|_{\rm FFE} \}$ have been well explored
both analytically and numerically in many previous studies \citep[see e.g.,][]{BZ77,Tanabe08, Contop13, Pan15, Pan15b}.
Similar to the FFE case, solving the MHD GS equation (\ref{GS_MHD}) is also eigenvalue problem, where $\Omega$ and $4\pi\eta L$ are the to-be-determined eigenvalues ensuring field lines smoothly cross the two Alfven surfaces \citep{Contop13, Nathan14, Pan17,Mahlmann18}.

\section{A split monopole example} \label{sec:eg}

As previewed in the Introduction, we aim to construct a framework for investigating MHD jet structure of spinning BHs, 
in which the EM fields $(F_{\mu\nu})$ and the fluid motion $(n, u^\mu)$ are self-consistently obtained 
given a proper plasma loading function $\eta(r,\theta)$ and proper boundary conditions.

In this section, we detail the procedure of consistently solving the two governing equations for an example of
the split monopole magnetic field configuration around a rapidly spinning central BH with a dimensionless spin $a=0.95$. For simplicity, we explore two different scenarios
with magnetization parameters $\sigma_*^{\rm out}=2\sigma_*^{\rm in}$ and $\sigma_*^{\rm out}=\sigma_*^{\rm in}$,
respectively. Remember that the loading function $\eta(r,\theta)$ is completely determined by the 
magnetization parameters via the definition (\ref{sigM}).

Boundary conditions used here are similar to those of force-free solutions. 
Explicitly, we choose $\Psi|_{\mu=0}=\Psi_{\rm max}$  on the equatorial plane, 
$\Psi|_{\mu=1}=0$ in the polar direction,  $\Psi_{,r}|_{r=r_{\rm H}}=0$ and $\Psi_{,r}|_{r=\infty}=0$
for the inner and outer boundaries, respectively. Here $r_{\rm H}$ is the radius of the event horizon.

\subsection{Numerical Techniques} \label{sec:tech}
We define a new radial coordinate $R = r/(r+1)$,  confine our 
computation domain $R\times\mu$ in the region $[R(r_{\rm H}), R_{\rm max}]\times[0,1]$,
and implement a uniform $256\times64$ grid. 
In practice, we choose $R_{\rm max}=0.995$, i.e., $r_{\rm max}\approx 200 M$.

The Bernoulli equation (\ref{eq:Bern}) and the MHD GS equation (\ref{GS_MHD}), governing the flow along the field lines
and field line configuration, respectively, are coupled. So we solve them one by one in an iterative way:
\begin{eqnarray} \label{Eq_set}
\left\{\begin{array}{c}
    \mathcal{L}\Psi^{(l)}=(\mathcal{S}_{\rm EM}+\mathcal{S}_{\rm MT})\{(\eta L)^{(l)}, n^{(l-1)}, u^{(l-1)}\}\ , \\
    \mathcal{F}\{u^{(l)}; (E/L)^{(l)}, \Omega^{(l)}, \Psi^{(l)}\} =0 \ ,
\end{array}
\right.
\end{eqnarray}
with $l=1,2,3,\cdots$ . In a given loop $l$, we solve the GS equation updating $\Psi$ and $\{\Omega, (\eta L)\}$ (with $\{n, u^\mu\}$ inherited from the previous loop  $l-1$), ensuring field lines smoothly cross the two \Alfven surfaces; 
in a similar way, we solve the Bernoulli equation updating $u^\mu$ and $(E/L)$ (with freshly updated $\Omega$ and $\Psi$ from solving the GS equation), ensuring a super-sonic inflow solution and an outflow solution satisfying the match condition 
(\ref{eq:EOLout}). Combing solutions to both equations and definitions of $\{\eta, E, L\}$, we finally obtain all the desired quantities $\{F_{\mu\nu}, n, u^\mu\}$ as functions of coordinates $r$ and $\theta$. 

We activate the iteration with an initial guess
\begin{eqnarray} \label{init}
\left\{\begin{array}{cccc}
    \Psi^{(0)}(r,\theta)&=&\Psi_{\rm max}(1-\cos\theta)\ ,  \\
    \Omega^{(0)}(\Psi)&=&0.5\Omega_{\rm H}\ ,  \\
    (\eta L)^{(0)}(\Psi)&=&\Omega_{\rm H} \Psi[2-(\Psi/\Psi_{\rm max})]/(8\pi)\ ,\\
    n^{(0)}(r,\theta) &=& u^{(0)}(r,\theta) = 0\ ,
\end{array}
\right.
\end{eqnarray}
where $\Omega_{\rm H}\equiv a/(r_{\rm H}^2+a^2)$ is the BH angular velocity.

The numerical techniques for tackling the two eigenvalue problems are detailed as follows:

\begin{itemize}
\item[\it Step 1] 
The MHD GS equation is a second-order differential
equation which degrades to first order on the
\Alfven surfaces where $\mathcal{A}(r,\theta)=0$. 
Numerical techniques for dealing this problem have been well developed
in previous force-free studies \citep{Contop13, Nathan14, Huang16,Huang18, Pan17,Mahlmann18}, and we briefly recap them here.

In each loop $l$, we solve the GS equation (\ref{GS_MHD}) with
the approximate solution obtained from the previous loop $\left\{\Omega^{(l-1)},(\eta L)^{(l-1)},\Psi^{(l-1)}\right\}$
as the initial guess. 
We evolve the flux function $\Psi^{(l)}$ using the overrelaxation technique with Chebyshev acceleration \citep{Press86}, and $\Psi^{(l)}(r,\theta)$ is updated on grid points except those in the vicinity of the two \Alfven surfaces. The  flux function $\Psi^{(l)}(r,\theta)$ on the \Alfven surfaces are obtained via interpolation from neighborhood grid points and the directional derivatives on the \Alfven surfaces \citep{Pan17}. 
Usually we obtain two different flux function $\Psi(r_{\rm A}^-)$ versus $\Psi(r_{\rm A}^+)$ 
on the \Alfven surface via interpolations from grid points inside and outside, respectively. 
To decrease this discontinuity, we adjust $\Omega^{(l)}(\Psi)$ at the outer Alfv\'en (OA) surface:
\begin{eqnarray}
	\Omega^{(l)}_{\rm new}(\Psi_{\rm new})&=& \Omega^{(l)}_{\rm old}(\Psi_{\rm old}) \nonumber\\
	&\ & + 0.05 [\Psi(r_{\rm OA}^+)-\Psi(r_{\rm OA}^-)],
\end{eqnarray}
with $\Psi_{\rm new}=0.5[\Psi(r_{\rm OA}^+)+\Psi(r_{\rm OA}^-)]$, 
where the subscript old/new represents quantities before/after the above adjustment;
and adjust both $\Omega^{(l)}(\Psi)$ and $(\eta L)^{(l)}(\Psi)$ at the inner Alfv\'en surface (IA):
\begin{eqnarray}
	\Omega^{(l)}_{\rm new}(\Psi_{\rm new})&=& \Omega_{\rm old}(\Psi_{\rm old}) \nonumber\\
	&\ & + 0.05[\Psi(r_{\rm IA}^+)-\Psi(r_{\rm IA}^-)],\nonumber\\
	(\eta L)^{(l)}_{\rm new}(\Psi_{\rm new})&=& (\eta L)^{(l)}_{\rm old}(\Psi_{\rm old}) \nonumber\\
	&\ & - 0.05[\Psi(r_{\rm IA}^+)-\Psi(r_{\rm IA}^-)],
\end{eqnarray}
with $\Psi_{\rm new}= 0.5[\Psi(r_{\rm IA}^+)+\Psi(r_{\rm IA}^-)]$.

After sufficient evolution, we obtain a converged solution $\{\Omega^{(l)}, (\eta L)^{(l)}, \Psi^{(l)}\}$ 
 which ensures field lines smoothly cross the two \Alfven surfaces.

\item[\it Step 2] The Bernoulli equation in the form of Eq.(\ref{Bern2}) is  a fourth-order polynomial equation in $u_p$
\citep{Camen86a,Camen86b,Camen87,Takahashi90,Fendt01,Fendt04,Levinson06,Pu15},
where the FM point is a standard `X'-type singularity, while the Alfv\'en point turns out to be a higher-order
singularity \citep{Weber67}.
Mathematically, a FM point is the location of a multiple root to the Bernoulli equation. 
The existence of FM point is very sensitive to the value of $(E/L)_{\rm in}^{(l)}$. 
For a slightly small value, there exists only sub-sonic solutions in the region $r<r_A^{\rm in}$, 
i.e., no multiple root. For a slightly larger value, on the other hand, there exists no global solution
extending from $r_{\rm H}$ to $r_A^{\rm in}$. Only for some specific choice, there exist a global super-sonic solution that crossing the FM point (see Fig.~\ref{fig:up} for numerical examples). 
We adjust $(E/L)_{\rm in}^{(l)}$ on each field line until an inflow solution that smoothly crosses the FM point is found. 

With the inflow solution in hand, $(E/L)_{\rm out}^{(l)}$ on each field line is uniquely determined by the match condition Eq.(\ref{eq:EOLout}), then it is straightforward to compute the outflow velocity.

\item[\it Step 3] Combining the inflow and the outflow solutions from Step 2,
the global fluid velocity $u^{(l)}$ and therefore the number density $n^{(l)},$ and the  Mach number $\mathcal{M}^{(l)}$ are obtained along each field line. We feed these quantities into the GS equation (\ref{GS_MHD})  for the next loop.
We iterate Step 1 to Step 3 until all quantities converge to a given precision.
\end{itemize}

We should point out that there is  an unphysical divergence 
arising from the idealized plasma loading on a single surface adopted in this paper. 
The particle number flux function $\eta$ is negative/positive for inflow/outflow and is not continous
on the loading surface. According to the 
definition of $\eta$ in Eq.(\ref{eq:eta}),
the particle number density $n$ is proportional to $\eta/u_p$ and therefore
diverges on the stagnation surface where $u_p=0$. The particle number density $n$ show up in 
the source terms of the MHD GS equation (\ref{GS_MHD}). To overcome the unphysical infinity,
we smooth them in the vicinity of the stagnation surface before feeding into Eq.(\ref{GS_MHD}).\footnote{
In addition, we usually obtain two different angular momentum flux per magnetic flux $(\eta L)_{\rm Bern}$ from the Bernoulli equation and $(\eta L)_{\rm GS}$ from the GS equation, respectively. The former is obtained from $E/L$ and Eq.(\ref{eq:Estag}), while the latter is one of the eigenvalues of the GS equation (\ref{GS_MHD}). In general, the two do not match exactly,  where $(\eta L)_{\rm GS}$ does approach $(\eta L)|_{\rm FFE}$ in the limit of $\sigma_*\rightarrow \infty$, while  $(\eta L)_{\rm Bern}$ does not. For the cases investigated in 
Section \ref{sec:results}, we find $(\eta L)_{\rm GS}$ is different from $(\eta L)_{\rm Bern}$ by 
$\lesssim 25\%$. This mismatch indicates that the particle number flux per magnetic 
flux $\eta(r,\theta)$ cannot be arbitrarily given, or perfectly conducting fluid is not a sufficient description here. }

\section{Numerical Results} \label{sec:results}

\subsection{Case Study}

In this subsection, we explore two nearly force-free cases with $\sigma_*\gg 1$: {\it Case 1} with magnetization parameters $\sigma_*^{\rm out} =\sigma_*^{\rm in}=75$  and {\it Case 2} with $\sigma_*^{\rm out} =2\sigma_*^{\rm in}=60$.
\begin{figure*}[h!]
\centering
\includegraphics[scale=0.9]{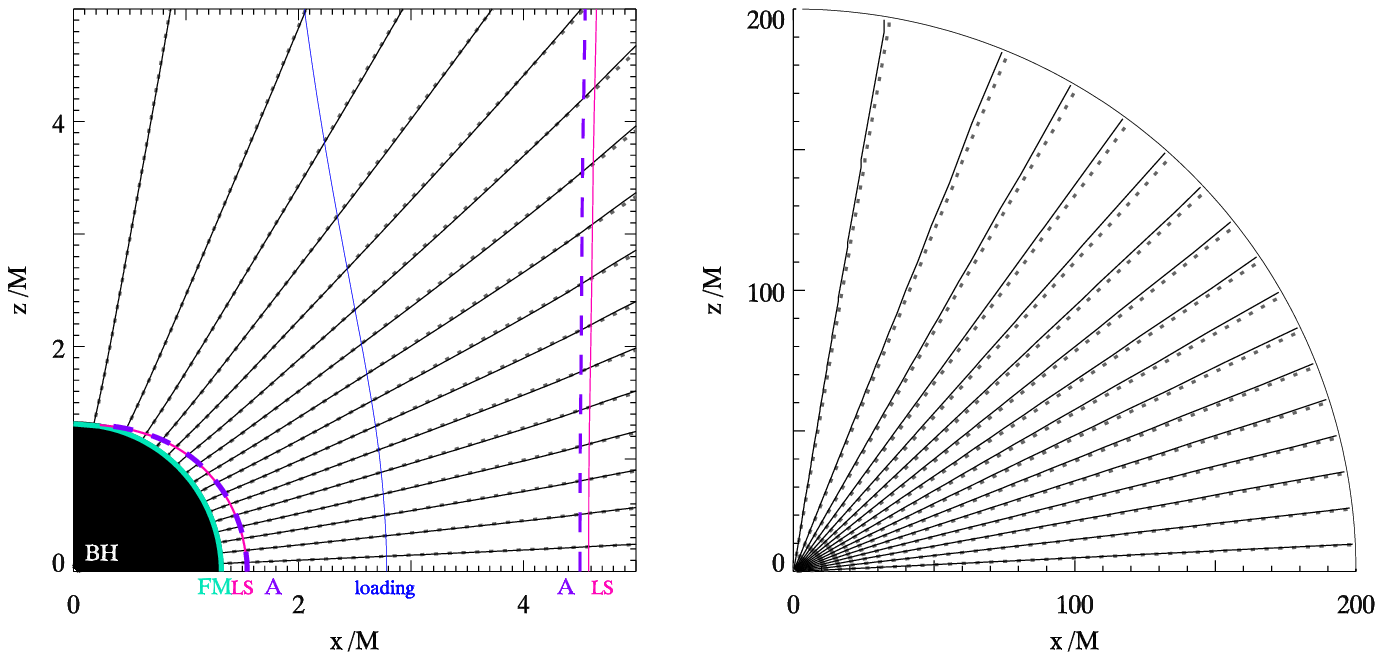}\\
\includegraphics[scale=0.9]{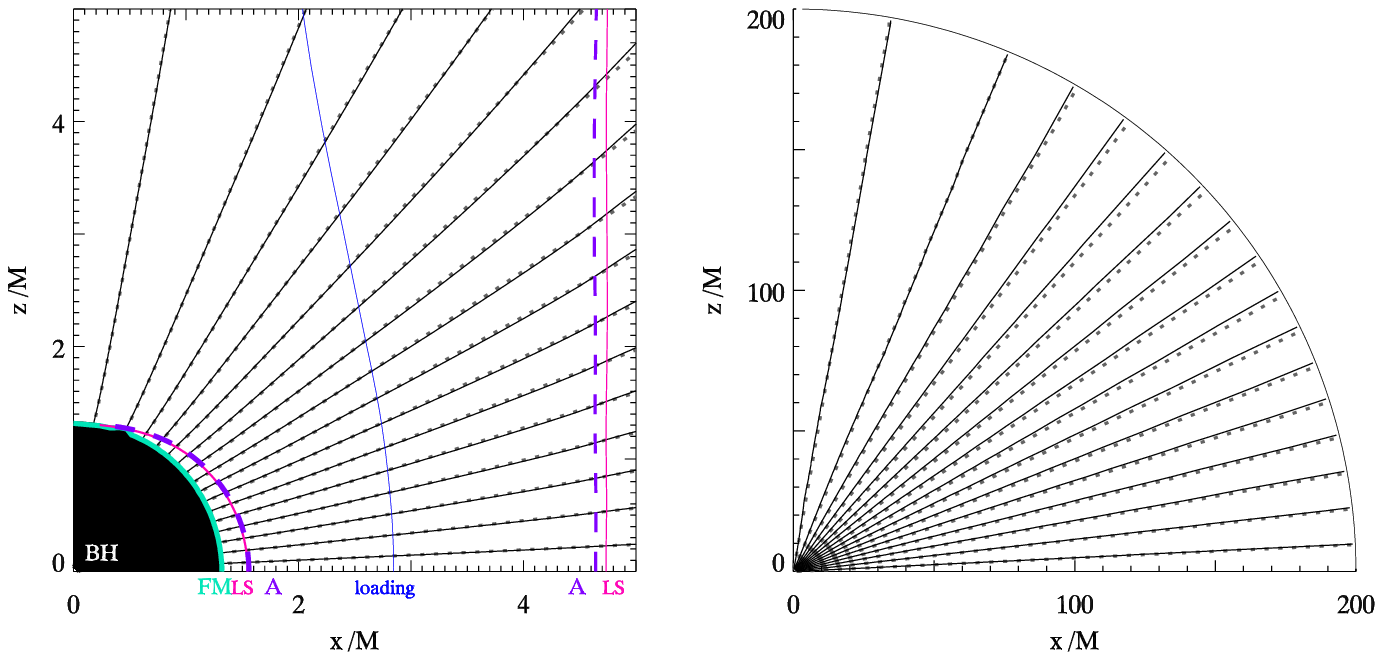}
\caption{{\it Top-left:} Comparison of the poloidal field line configuration of an MHD jet of {\it Case 1} (black solid lines)
with parameters $\sigma_*^{\rm out} = \sigma_*^{\rm in}=75$
versus its FFE counterpart (black dotted lines).
The loading surface (loading), Alfv\'en surfaces (A), light surfaces (LS),
and inner fast magnetosonic (FM) are presented in blue, dashed purple,
magenta, and aqua line, respectively.
The configuration of a force-free jet is also shown for comparison (dotted line).
{\it Top-right:} A zoom-out configuration of the left panel.
{\it Bottom:} The poloidal field line configuration of an MHD jet solution of {\it Case 2}
with parameters $\sigma_*^{\rm out}=2\sigma_*^{\rm in}=60$. \label{fig:fl}}
\end{figure*}

\begin{figure*}[h!]
\centering
\includegraphics[scale=0.9]{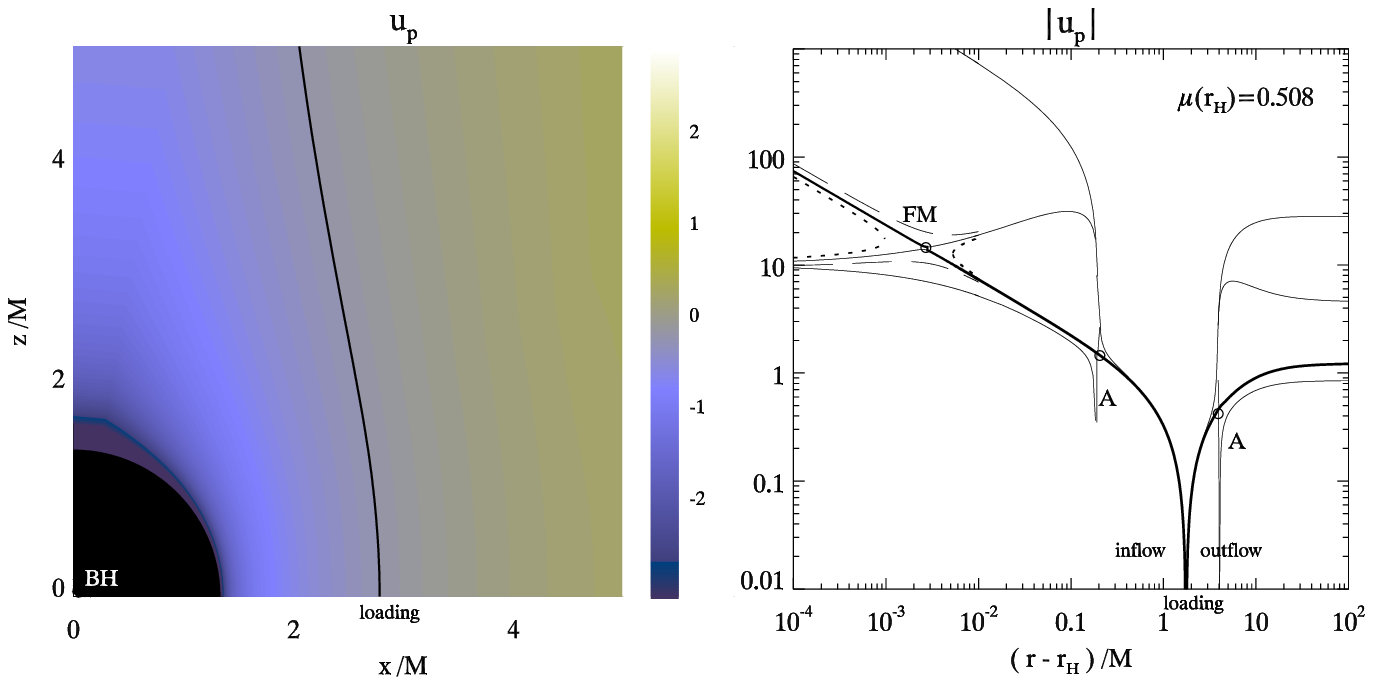}\\
\includegraphics[scale=0.9]{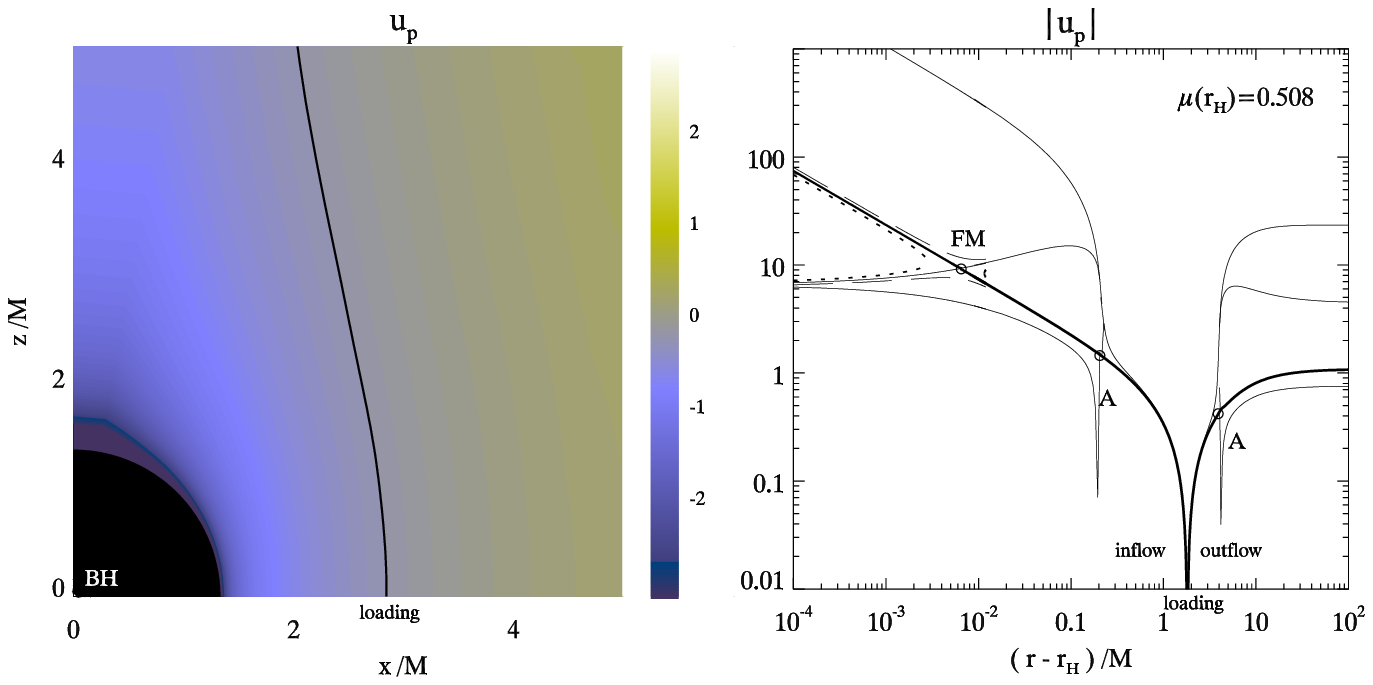}
\caption{{\it Top-left:} The poloidal fluid velocity $u_p$ of {\it Case 1}.
{\it Top-right:} $|u_p|$ as a function of $r$ on the field line with foot-point on the event horizon $\mu(r_{\rm H})=0.508$, where the solutions with correct eigenvalue $E/L$ are shown in solid lines and those with $E/L$ of slightly larger/smaller values are shown in dotted/long-dashed lines.
The Alfv\'en points and the FM point are marked by open circles.
{\it Bottom:} The fluid motion configuration of {\it Case 2},
similar to top panels.  \label{fig:up}}
\end{figure*}

\begin{figure*}[h!]
\centering
\includegraphics[scale=0.9]{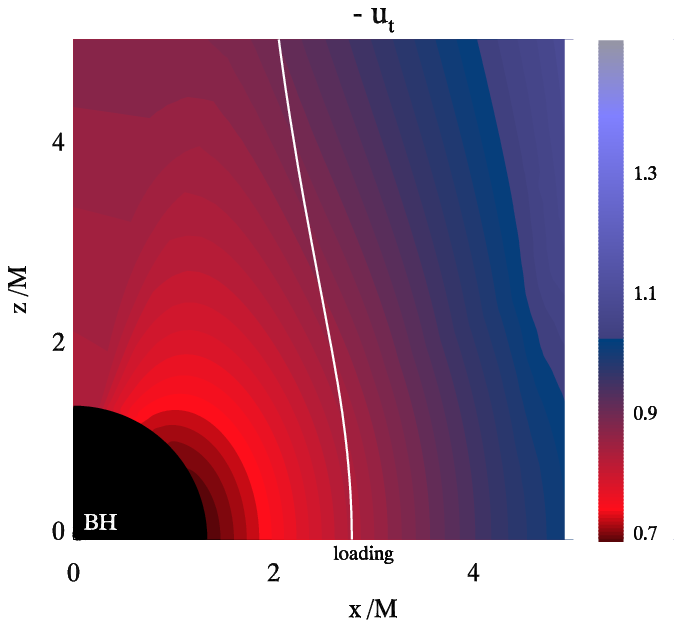}
\includegraphics[scale=0.9]{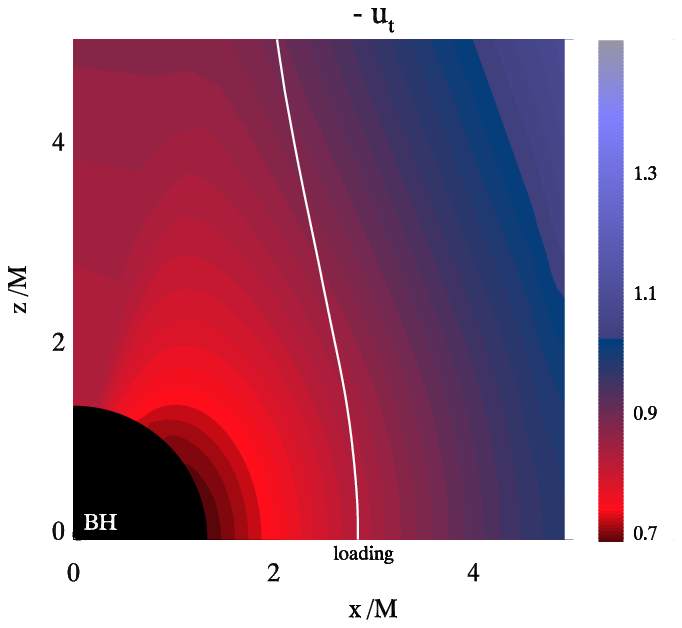}
\caption{The configuration of particle energy $-u_t$ of {\it Case 1} ({\it left}) 
and {\it Case 2} ({\it right}). \label{fig:ut}}
\end{figure*}

In Fig.~\ref{fig:fl},  we compare the magnetic field line configurations of MHD solutions with that of FFE version. For both cases, which are nearly force-free, the deviation from the FFE version is as expected small, though the deviation of {\it Case 1} from the FFE solution is more obvious than that of {\it Case 2},
due to more efficient outflow acceleration in the former case (see Fig.~\ref{fig:ut}).

In Fig.~\ref{fig:up}, we show the poloidal velocity $u_p$. To clarify the different singularity types of FM points and \Alfven points, we explicitly show all the solutions (both physical one and unphysical ones) to the Bernoulli equation
(\ref{Bern2}) for the field line with foot point  $\mu(r_{\rm H})=0.508$, where the solutions with correct eigenvalue $E/L$ are shown in solid lines and the solutions with $E/L$ slightly off the correct value are shown in dotted and dashed lines. 
At the FM point which is a `X'-type singularity, there exists a multiple root of the Bernoulli equation for correct eigenvalue $E/L$, while no global solution (dotted lines) or no multiple root exists (long-dashed lines)  for $E/L$ of slightly off value.
At the \Alfven point which is a higher-order singularity, there exists multiple root  no matter $E/L$ takes the correct eigenvalue or not. For the physical solution, the inflow passes along $r_*\to r_{\rm A}\to r_{\rm FM}\to r_{\rm H}$ and the outflow passes along $r_*\to r_{\rm A}\to\infty$.

\begin{figure*}[h!]
\centering
\includegraphics[scale=0.9]{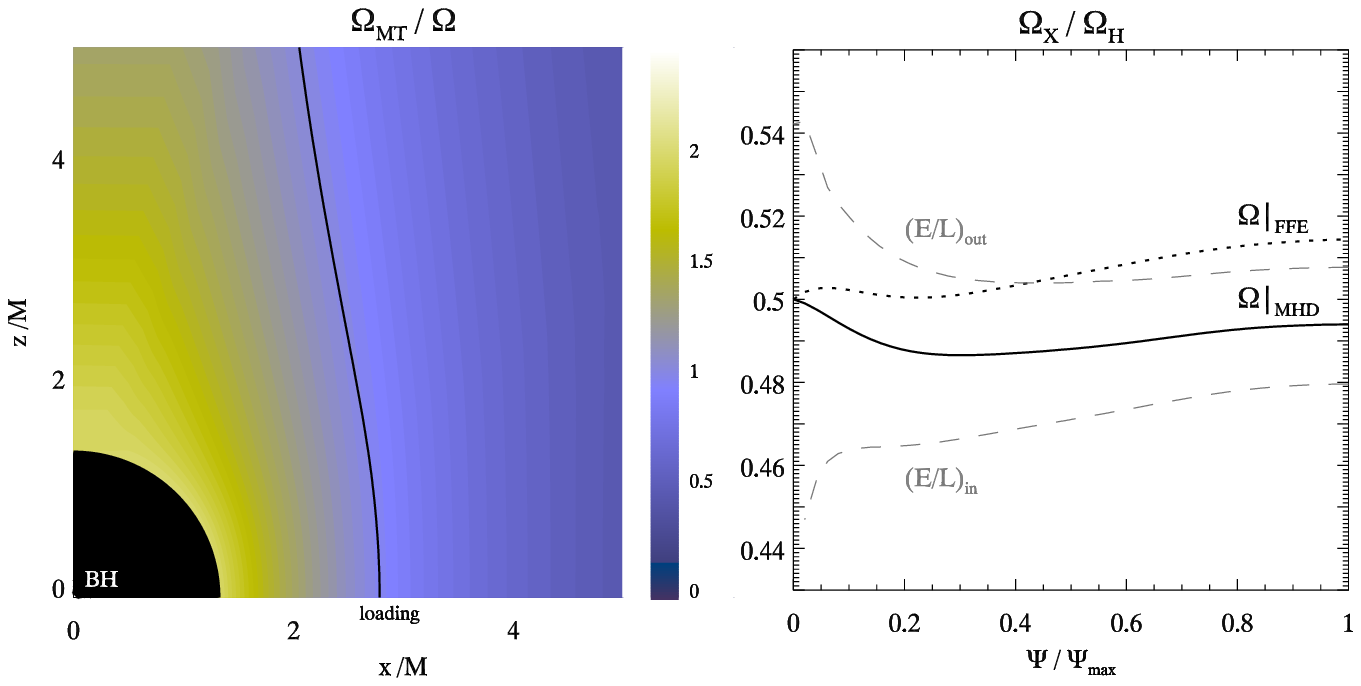}\\
\includegraphics[scale=0.9]{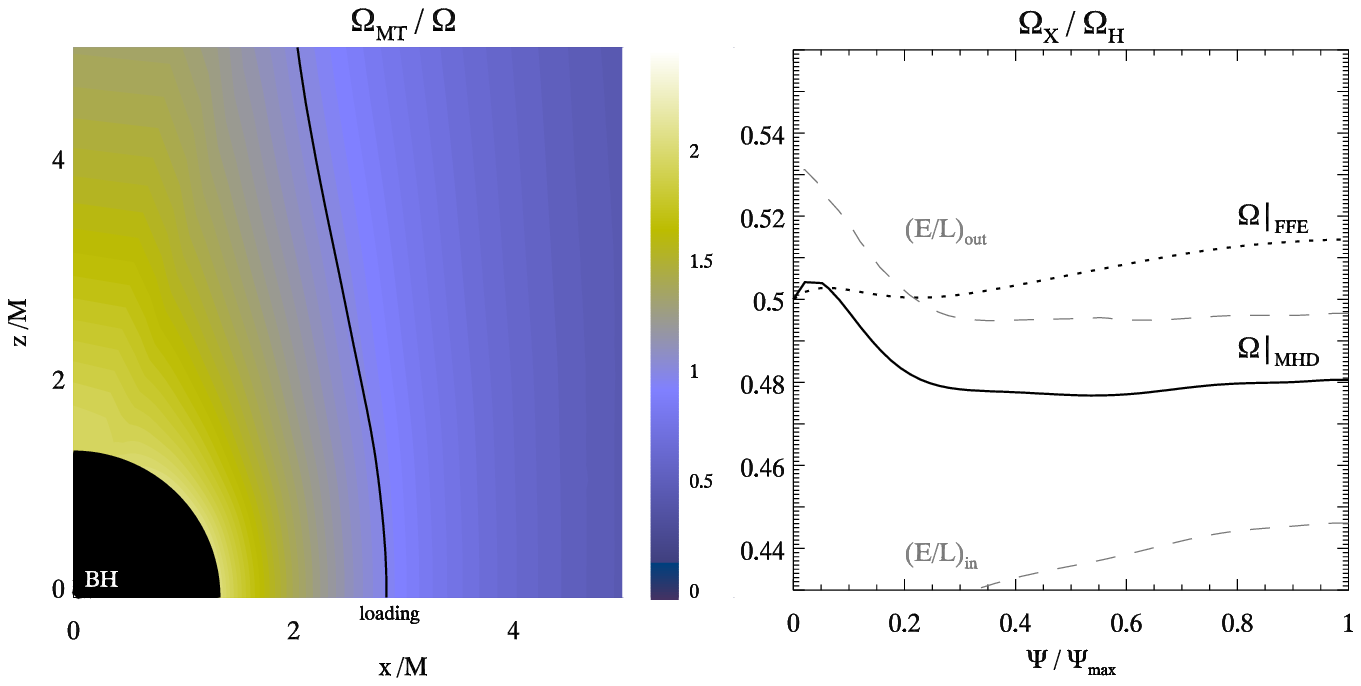}
\caption{{\it Top-left:} The fluid angular velocity $\Omega_{\rm MT}$ of {\it Case 1}.
{\it Top-right:} Comparison of a few angular velocity like quantities $\{\Omega|_{\rm FFE}, \Omega|_{\rm MHD}, (E/L)_{\rm in;out}\}$ of {\it Case 1} .
{\it Bottom Panels:}  same as the top ones except for {\it Case 2}.  \label{fig:Om}}
\end{figure*}

In Fig.~\ref{fig:Om}, we show the fluid angular velocity $\Omega_{\rm MT}$, the angular velocity of magnetic
field lines $\Omega$ (solid lines), and the eigvenvalues $(E/L)_{\rm in;out}$ (dashed grey lines). 
Consistent with our intuition, we find the rotation of magnetic field lines is dragged down by the fluid inertia compared with the FFE case, i.e., $\Omega|_{\rm MHD} < \Omega|_{\rm FFE}$. Due to nonzero poloidal velocity of both
inflow and outflow, the fluid does not corotate with the field lines, i.e., $\Omega_{\rm MT}<\Omega$ for inflow  and $\Omega_{\rm MT}>\Omega$ for outflow. Specifically, the fluid angular velocity on the
event horizon $\Omega_{\rm MT}(r=r_{\rm H})$ slightly exceeds the BH angular velocity $\Omega_{\rm H}$,
which guarantees the fluid energy to be positive on the horizon (see Fig.~\ref{fig:ut}). 

In Fig.~\ref{fig:ut}, we show the specific particle energy $-u_t$ for the two cases.
Both of them are positive everywhere, while the outflow of \emph{Case 1} gains more efficient
acceleration.

\subsection{Energy Extraction Rates}

\begin{figure}[h!]
\centering
\includegraphics[scale=1]{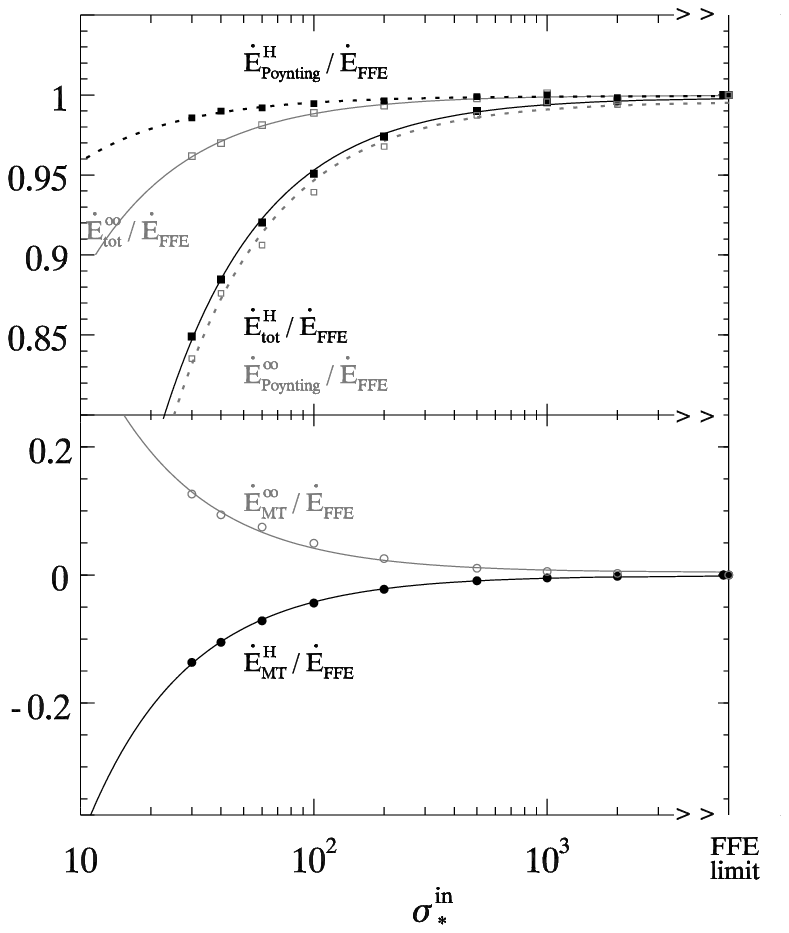}
\caption{Results of the energy extraction rates in relation to $\sigma_*^{\rm in}$ with $\sigma_*^{\rm out}=2\sigma_*^{\rm in}$ assumed.
The energy rates measured at the event horizon
$\{ \dot{E}_{\rm tot}^{\rm H}, \dot{E}_{\rm Poynting}^{\rm H}, \dot{E}_{\rm MT}^{\rm H}\}$, are presented in filled squares, small filled squares, and filled circles, respectively. The solid black line in top panel, the dotted black line in top panel, and the solid black line in bottom panel are the corresponding fitting curves.
Similarly, the energy rates measured at infinity
$\{ \dot{E}_{\rm tot}^{\infty}, \dot{E}_{\rm Poynting}^{\infty}, \dot{E}_{\rm MT}^{\infty}\}$, are presented in open symbols and the solid grey lines are the corresponding fitting curves. \label{fig:Edot}}
\end{figure}

In this subsection, we investigate the energy extraction rate from the central BH via the MHD jet,
which is defined as 
\be 
\begin{aligned}
      \dot{E}_{\rm tot}(r)
      &= -2\pi \int_0^{\pi} \sqrt{-g} T^r_{\ t}(r) {\rm d} \theta\ , \\ 
      &= 4\pi\int_0^{\Psi_{\rm max}} (\eta E)(r) {\rm d}\Psi \ , \\ 
      &= 4\pi\int_0^{\Psi_{\rm max}} [(E/L)\times(\eta L)](r) {\rm d}\Psi \ ,
\end{aligned}
\ee 
where we have used Eqs.(\ref{eq:eta}-\ref{eq:EandL}) in the second line. In the third line, $E/L$ and $\eta L$ are the eigenvalues of the Bernoulli equation (\ref{eq:Bern}) and of the GS equation (\ref{GS_MHD}), respectively. In the similar way, we can define its matter/electromagnetic component as
\begin{equation} \label{eq:Edot}
\begin{aligned}
	\dot{E}_{\rm MT}(r)&=  4\pi\int_0^{\Psi_{\rm max}}(-  \eta m u_t)(r) {\rm d}\Psi\ , \\
	\dot{E}_{\rm Poynting}(r)
	&= 4\pi \int_0^{\Psi_{\rm max}}(\Omega I/4\pi)(r) {\rm d}\Psi \ .\\
	&= \dot{E}_{\rm tot}(r) - \dot{E}_{\rm MT}(r)\ . 
\end{aligned}
\end{equation}

We measure these energy extraction rates at $r=r_{\rm H}$/ $r=\infty$, and quantify their dependence
on the magnetization parameter $\sigma_*$. In practice, we find that these energy extraction rates are not sensitive to the value of $\sigma_*^{\rm out}$,
except the matter component of energy rate at infinity $\dot{E}_{\rm MT}^\infty$. 
Without loss of generality, we only show the rates in relation to $\sigma_*^{\rm in}$
for the $\sigma_*^{\rm out}=2\sigma_*^{\rm in}$ scenario in Fig.~\ref{fig:Edot}, where
all the rates are displayed in unit of the energy extraction rate in the force-free limit 
$\dot E_{\rm FFE}\approx 0.4 (\Psi_{\rm max}^2/4\pi)$.

As we see in Fig.~\ref{fig:Om}, the rotation of magnetic field lines is dragged down by
the loaded plasma, i.e., $\Omega|_{\rm MHD} < \Omega|_{\rm FFE}$, while the fluid that does not corotate with
the field lines with angular velocity $\Omega_{\rm MT}|_{\rm inflow} > \Omega > \Omega_{\rm MT}|_{\rm outflow}$, tends to bend the field lines and induce a stronger $\phi$-component of magnetic field, i.e., $I|_{\rm MHD} > I|_{\rm FFE}$. The net result is that the Poynting energy extraction rate on the event horizon $\dot E_{\rm Poynting}^{\rm H}$ has little dependence on the magnetization. Going outward along the field lines, part of the Poynting flux is converted into the fluid kinetic energy. For the case with magnetization parameter $\sigma_*^{\rm in} = 30$, the matter component makes up about $13\%$ of the total energy flux at infinity.

\subsection{Penrose Process}\label{subsec:Penrose}

\begin{figure}[h!]
\centering
\includegraphics[scale=0.68]{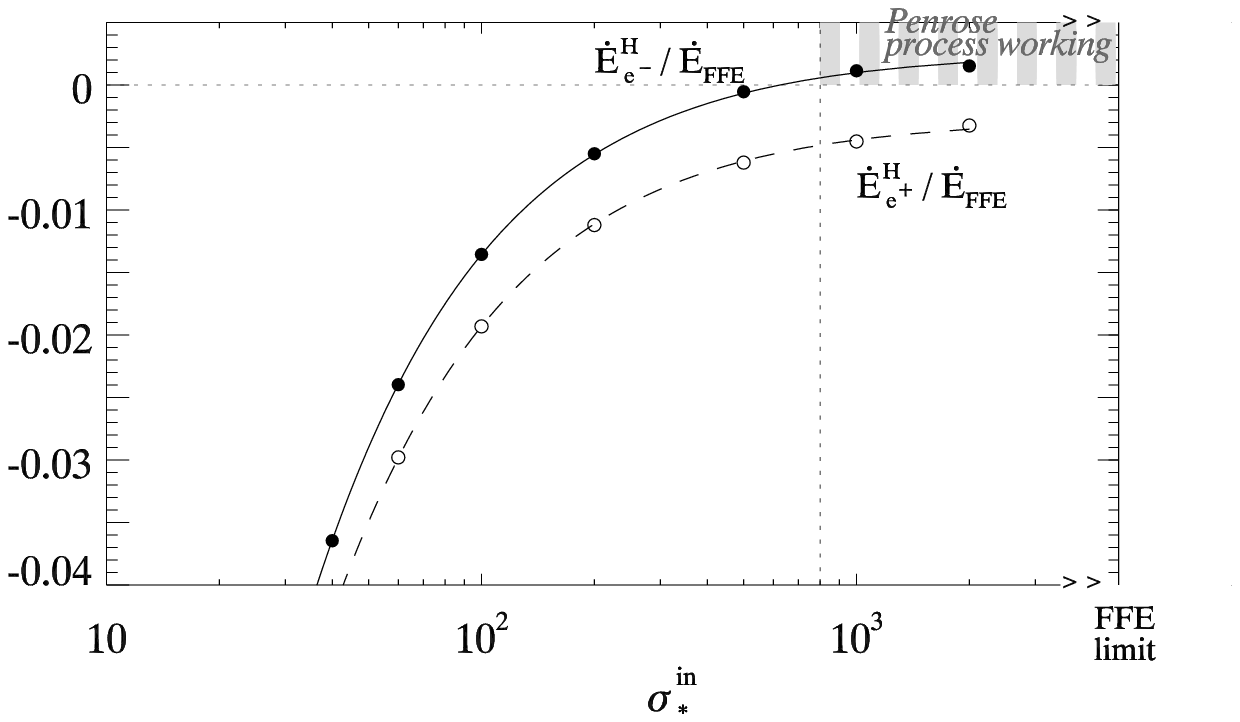}
\caption{The positron/electron component of energy extraction rates in relation to $\sigma_*^{\rm in}$. 
The energy rates  $\{\dot{E}^{\rm H}_{e^+}, \dot{E}^{\rm H}_{e^-}\}$ 
are presented in open and filled circles, respectively. 
The solid and dashed lines are the corresponding fitting curves.
The shaded regime denotes where the Penrose process is 
working (for the electron component). \label{fig:EdMT}}
\end{figure}

An implicit assumption in our MHD jet model is the two-fluid description, since 
the electric current density $j^\mu$ is \emph{not} proportional to the fluid velocity $u^\mu$.
Therefore we can decompose the charged fluid as two oppositely charged components, 
positron ($e^+$) and electron ($e^-$). \footnote{Though there is a degree of freedom in doing 
this decomposition, e.g., we can also decompose the fluid as electrons and ions, it does not change
our conclusion qualitatively.} We denote the number densities and the velocity fields as $n_\pm$ 
and $u^\mu_\pm$, respectively, which are related to $j^\mu$ and $nu^\mu$ via relations
\begin{eqnarray}
    j^\mu&=& e(n_+u^\mu_+ - n_-u^\mu_-)\nonumber\\
    mnu^\mu&=& m(n_+u^\mu_+ + n_-u^\mu_-).
\end{eqnarray}
Consequently, we obtain 
\be 
 m(n u^\mu)_\pm =  \frac{1}{2}[\pm j^\mu(m/e) + nmu^\mu] \ ,
\ee 
and we can decompose the matter energy flux into two components $\dot{E}_{e^\pm}$.
Here we are only interested in the energy extraction rates on the event horizon
\begin{eqnarray}
    \dot{E}^{\rm H}_{e^+}&=& 4\pi \int_0^{\Psi_{\rm max}} \frac{(-\eta m n_+u_{t+})(r_{\rm H})}{n(r_{\rm H})} {\rm d}\Psi\ ,\nonumber\\
    \dot{E}^{\rm H}_{e^-}&=& 4\pi \int_0^{\Psi_{\rm max}} \frac{(-\eta m n_-u_{t-})(r_{\rm H})}{n(r_{\rm H})} {\rm d}\Psi \ .
\end{eqnarray}
As an example, we choose the horizon enclosed magnetic flux $\Psi_{\rm max}=1000(m/e)$, and show  $\dot{E}^{\rm H}_{e^+}/\dot{E}_{\rm FFE}$ 
and $\dot{E}^{\rm H}_{e^-}/\dot{E}_{\rm FFE}$ in relation to $\sigma_*^{\rm in}$ in Fig.~\ref{fig:EdMT}. 
The energy extraction rate from positrons is always negative, while 
the energy extraction rate from electrons, become positive 
when the plasma loading is low enough. In this regime, denoted in shades in Fig.~\ref{fig:EdMT}, the magnetic Penrose process is working, though, only for one of the two charged component. \footnote{We should not do any quantitative interpretation for the results of this subsection, because the two-fluid decomposition done here is not accurate,  e.g., there is no guarantee for the velocity of each component $u^\mu_\pm$ to be timelike and normalized.
We will leave a more accurate two-fluid description of MHD jet structure \citep{Koide09, Liu18} to future work .} This finding is in good agreement with recent particle-in-cell simulations \citep{Parfrey18}.

\section{Summary and Discussion}\label{sec:summary}
\subsection{Summary}
To describe the MHD structure of BH jets, we need a minimum set of quantities as functions of spacetime: 
Maxwell tensor $F_{\mu\nu}$, fluid rest mass density $\rho$ (or equivalently particle number density $n$), and 
fluid four-velocity $u^\mu$. For determining all these quantities self-consistently, we constructed a full MHD framework, in which EM fields and fluid motion are governed by the MHD GS equation (\ref{GS_MHD}) and the Bernoulli equation (\ref{eq:Bern}), respectively. From these two governing equations, we can completely determine $\{F_{\mu\nu}, \rho ,u^\mu\}$ given proper boundary conditions and a proper plasma loading function $\eta(r,\theta)$ (see Eq.(\ref{eq:eta})). As an example, we consider a split monopole field configuration and idealized plasma loading on the stagnation surface.

Assuming steady and axisymmetric jet structure, and perfectly conductive plasma within the jet, the EM fields are 
completely determined by three functions: the magnetic flux $\Psi(r, \theta)$, the angular velocity of magnetic field 
lines $\Omega(\Psi)$ and the poloidal electric current $I(r,\theta)$ (see Eq.(\ref{eq:Maxwell})). 
Given fluid energy density $\rho$ and velocity $u^\mu$, the MHD GS equation (\ref{GS_MHD}) turns out to be a second-order differential equation with respect to $\Psi(r,\theta)$ which degrades to be first-order on the two \Alfven surfaces. Solving the GS equation is an eigenvalue problem, with eigenvalues $\Omega(\Psi)$
and $I(r,\theta)$ (or more precisely, the conserved quantity $4\pi\eta L(\Psi)$ defined in Eq.(\ref{eq:EandL})) to be determined ensuring field lines smoothly cross the \Alfven surfaces.

Given EM fields $F_{\mu\nu}$, the Bernoulli equation turns out to be a fourth-order polynomial equations in the poloidal fluid velocity $u_p$. Solving the Bernoulli equation is also an eigenvalue problem,  with the eigenvalue $(E/L)_{\rm in}$ to be determined ensuring the inflow smoothly cross the FM surface, and $(E/L)_{\rm out}$ to be determined by the match condition (\ref{eq:EOLout}) on the loading surface. With both $E/L$ and $u_p$ obtained, it is straightforward to obtain $n$ and $u^\mu$ via Eqs.(\ref{eq:eta},\ref{eq:EandL}) and the  normalization 
condition $u\cdot u=-1$. 

The two governing equations are coupled, therefore we numerically solved them in an iterative way (see Sec.~\ref{sec:tech}). 
As a result, we find the rotation of magnetic field lines is dragged down by the plasma loaded, i.e., $\Omega|_{\rm MHD} < \Omega|_{\rm FFE}$; for the fluid angular velocity, we find  $\Omega_{\rm MT}|_{\rm outflow}<\Omega<\Omega_{\rm MT}|_{\rm inflow}$ ; 
the non-corotating fluid tends to bend the field lines and induce a stronger $\phi$-component of magnetic field, therefore 
a stronger poloidal electric current
i.e., $I|_{\rm MHD} > I|_{\rm FFE}$. The net result is that the Poynting energy extraction on the horizon is insensitive to the
magnetization, i.e., $\dot E_{\rm Poynting}^{\rm H} |_{\rm MHD} \approx \dot E_{\rm Poynting}^{\rm H}|_{\rm FFE}$ (see Fig.~\ref{fig:Edot}).
Going outward along the field lines, part of the Poynting flux is converted to the fluid kinetic energy. For the case we explored
with $\sigma_*^{\rm in} = 30$, the matter component makes up $\sim 13\%$ of the total energy flux at infinity.

Finally, we examined the MHD Penrose process for the cases we numerically solved. We found the specific fluid energy $-m u_t$ is always positive on the event horizon, i.e., the MHD Penrose process is not working and therefore the BZ mechanism defines fully the jet energetics. However, if we decompose the charged fluid as two oppositely charged components ($e^\pm)$, we found the magnetic Penrose process does work for one of the two components when the plasma loading is low enough (see Fig.~\ref{fig:EdMT}).

\subsection{Discussion}

As a first step towards a full MHD jet model, we have investigated the MHD jet structure
of split monopole geometry assuming an idealized plasma loading on the stagnation  surface. 
This simplified plasma loading gives rise to a few unphysical problems in the vicinity of the loading surface, 
including divergence of particle number density $n(r_*)$, which shows up in the source terms of the MHD
GS equation (\ref{GS_MHD_S}). To avoid the singularity arising from the unphysical divergence, we smoothed the
function $n(r,\theta)$ in the vicinity of the loading surface. Another consequence of the simplified 
plasma loading is that we must impose the continuity equation (\ref{eq:EOLout}) to ensure the EM fields to be continuous across the loading surface. As a result,  $(E/L)_{\rm out}$ is specified by $(E/L)_{\rm in}$, i.e., $r_{A,\rm out}$ is specified by $r_{A, \rm in}$. Therefore, we lose the freedom to adjust $(E/L)_{\rm out}$ until a supersonic outflow solution is found
as we did for the inflow solution. Consequently, all the outflow solutions obtained in this paper are subsonic (see Fig.~\ref{fig:up}). 

In future work, we aim to investigate a full MHD jet model with a more realistic extending loading zone 
where the plasma injection is described by a continuous function $\eta(r,\theta)$. Then all the unphysical
discontinuity and divergence described above would be avoided. For the extending plasma loading, the smooth EM fields would be naturally preserved, and the continuity requirement would not be a constraint. As a result,
we can adjust $(E/L)_{\rm out}$ for finding a supersonic outflow solution, which is more consistent with recent observations \citep{Hada16,Mertens16}.
In addition to the plasma loading, the BH surroundings also play an important role in shaping the jet structure \citep[e.g.][]{Tchek10,Beskin17}. The role of more realistic BH environment, including accretion flows and hot plasma with non-zero pressure will also be considered in future work.

\section*{Acknowledgements}
We thank the referee for his/her careful reading of this manuscript and giving insightful suggestions.
L.H. thanks the support by the National
Natural Science Foundation of China (grants 11590784 and 11773054),
and Key Research Program of Frontier Sciences, CAS (grant No. QYZDJ-SSW-SLH057).
Z.P. thanks Hung-Yi Pu for his invaluable help throughout this research.  
Z.P. is  supported by Perimeter Institute for Theoretical Physics. Research at Perimeter Institute
is supported  by the Government of Canada through the Department of Innovation, 
Science and Economic Development Canada and by the Province of Ontario through the Ministry of Research, Innovation and Science.
C.Y. has been supported by the National Natural Science Foundation of China (grants 11521303, 11733010 and 11873103).
This work made extensive use of the NASA Astrophysics Data System and
of the {\tt astro-ph} preprint archive at {\tt arXiv.org}.

\appendix
\section{Derivation of $\{ D_\Psi^\parallel \eta, D_\Psi^\parallel(\eta E), D_\Psi^\parallel(\eta L) \}$}
\label{sec:D_der}

Using Eq.(\ref{eq:eta}), it is straightforward to see
\be
\begin{aligned}
D_\Psi^\parallel (\eta) 
&= \frac{1}{\sqrt{-g}}(\Psi_{,\theta}\partial_r -  \Psi_{,r}\partial_\theta)\eta \ , \\ 
&= \frac{1}{\sqrt{-g}}\left[\partial_r(\eta \Psi_{,\theta}) -  \partial_\theta(\eta\Psi_{,r}) \right]\ , \\ 
&=\frac{1}{\sqrt{-g}}\frac{\partial}{\partial x^A} (\sqrt{-g}n u^A) \ , \\
&= (nu^\mu)_{;\mu} \ ,
\end{aligned}
\ee
from which we conclude that $D_\Psi^\parallel \eta$ is the source function of particle number density; i.e.,
$D_\Psi^\parallel (\eta)$ vanishes outside the plasma loading zone and is positive inside.
Due to the existence of plasma loading, the energy conservation of the electromagnetic fields and plasma system is written as
$T^{\mu\nu}_{\ \ ;\nu} = S^\mu$ [Eq.~(\ref{eq:smu})], where the source term $S^\mu$ comes from plasma loading and in this paper we have assumed $S^\mu=(D_\Psi^\parallel \eta) m u^\mu$. As a result,
\be
\begin{aligned}
(\xi_\mu T^{\mu\nu})_{;\nu} 
	&= \xi_{\mu;\nu}   T^{\mu\nu} + \xi_\mu T^{\mu\nu}_{\ \ ;\nu} \\
	&= 0 + (D_\Psi^\parallel \eta) mu^\mu \xi_\mu \\
	&= (D_\Psi^\parallel \eta) mu_t \ ,
\end{aligned}
\ee
where $\xi = \partial_t$ is the timelike Killing vector.
On the other hand,
\be
\begin{aligned}
      (\xi_\mu T^{\mu\nu})_{;\nu} 
      &= (T^\nu_{\ t})_{;\nu} \\
	 &= \frac{1}{\sqrt{-g}}\frac{\partial}{\partial x^A} (\sqrt{-g} T^A_{\ \ t}) \\
      &= \frac{1}{\sqrt{-g}} \frac{\partial}{\partial r} \left(\eta m u_t \Psi_{,\theta} + \frac{1}{4\pi}\sqrt{-g} F^{r\theta}\Omega\Psi_{,\theta} \right) \\
      &\qquad + \frac{1}{\sqrt{-g}} \frac{\partial}{\partial \theta} \left(-\eta m u_t \Psi_{,r} - \frac{1}{4\pi}\sqrt{-g} F^{r\theta}\Omega\Psi_{,r} \right) \\
      &= \frac{1}{\sqrt{-g}}  (\Psi_{,\theta} \partial_r - \Psi_{,r} \partial_\theta) \left(\eta m u_t -\frac{\Omega I}{4\pi}\right) \\
      &=-D_\Psi^\parallel (\eta E) \ .
\end{aligned}
\ee
Therefore we arrive at $D_\Psi^\parallel (\eta E) = (D_\Psi^\parallel \eta)(-m u_t)$. In the similar way,
we can derive $D_\Psi^\parallel (\eta L) = (D_\Psi^\parallel \eta)(m u_\phi)$.

\section{Derivation of the MHD-GS equation (\ref{GS_MHD})}
\label{sec:GS_der}

We now expand Eq.(\ref{eqn_MHD}) in terms of $\Psi(r,\theta), \Omega(\Psi), \eta(\Psi), \mathcal{M}(r,\theta), u_t(r,\theta), u_\phi(r,\theta)$, where the electromagnetic force have been derived in many previous FFE studies

\be
\begin{aligned}
\label{GS_EM}
	- \frac{F^A_{\ \phi}}{F_{C\phi}F^C_{\ \phi}} (F_{A\nu}j^\nu) = - \frac{1}{4\pi\Sigma \sin^2\theta} \times
	&\ \left\{\ \left[\Psi_{,rr} + \frac{\sin^2\theta}{\Delta} \Psi_{,\mu\mu} \right]\ \mathcal{K}(r,\theta;\Omega)  \right. \\
	&\ +\left[ \Psi_{,r} \partial^\Omega_r  + \frac{\sin^2\theta}{\Delta} \Psi_{,\mu} \partial^\Omega_\mu \right] \ \mathcal{K}(r,\theta;\Omega)  \\
	&\ +\frac{1}{2} \left[ (\Psi_{,r})^2 + \frac{\sin^2\theta}{\Delta} (\Psi_{,\mu})^2 \right] D_\Psi^\perp\Omega\ \partial_\Omega \mathcal{K}(r,\theta;\Omega) \\ 
	&\ -\left. \frac{\Sigma}{\Delta} I D_\Psi^\perp I \ \right\}\ , 
\end{aligned}
\ee
where $\mathcal{K}(r,\theta;\Omega) = -k_0$. With the aid of the normalization condition $u^\mu u_\mu=-1$ and $(u^\mu u_{\mu})_{;A}=0$, the matter acceleration is rewritten as
\be
\begin{aligned}
\label{eqn_MT}
	\frac{F^A_{\ \phi}}{F_{C\phi} F^C_{\ \phi}} ( mn u^\nu u_{A;\nu} ) 
	&= - mn \left( u^t D^\perp_\Psi u_t + u^\phi D^\perp_\Psi u_\phi \right)  - \frac{m\eta}{\sqrt{-g}}(u_{r,\theta}-u_{\theta,r})\ \\
	&= - \frac{1}{4\pi\Sigma\sin^2\theta} \times \ \left\{\ \left[\Psi_{,rr} + \frac{\sin^2\theta}{\Delta} \Psi_{,\mu\mu} \right]\ \mathcal{M}^2(r,\theta)  \right.  \\
	&\ \qquad\qquad\qquad\qquad + \left[ \Psi_{,r} \partial_r  + \frac{\sin^2\theta}{\Delta} \Psi_{,\mu} \partial_\mu \right]\ \mathcal{M}^2(r,\theta)  \\
	&\ \qquad\qquad\qquad\qquad - \left[ (\Psi_{,r})^2 + \frac{\sin^2\theta}{\Delta} (\Psi_{,\mu})^2 \right] \frac{D^\perp_\Psi\eta}{\eta}\ \mathcal{M}^2(r,\theta)  \\
	&\ \qquad\qquad\qquad\qquad \left. +\ 4\pi\Sigma\sin^2\theta mn \left( u^t D^\perp_\Psi u_t + u^\phi D^\perp_\Psi u_\phi \right)\ \right\}\ . 
\end{aligned}
\ee

\end{document}